\documentclass[amssymb,aps,prl,twocolumn,groupedaddress,showpacs]{revtex4}

\usepackage{graphicx}
\usepackage{psfrag}

\begin{document}

\title{Layering transitions for adsorbing polymers in poor solvents}
\author{J. Krawczyk}
\email{krawczyk.jaroslaw@tu-clausthal.de}
\affiliation{Institut f\"ur Theoretische Physik, Technische Universit\"at Clausthal, Arnold Sommerfeld Stra\ss e 6, D-38678 Clausthal-Zellerfeld, Germany}
\author{A. L. Owczarek}
\email{aleks@ms.unimelb.edu.au}
\affiliation{Department of Mathematics and Statistics, The University of Melbourne, 3010, Australia}
\author{T. Prellberg}
\email{thomas.prellberg@tu-clausthal.de}
\affiliation{Institut f\"ur Theoretische Physik, Technische Universit\"at Clausthal, Arnold Sommerfeld Stra\ss e 6, D-38678 Clausthal-Zellerfeld, Germany}
\author{A. Rechnitzer}
\email{andrewr@ms.unimelb.edu.au}
\affiliation{Department of Mathematics and Statistics, The University of Melbourne, 3010, Australia}

\date{\today }

\begin{abstract} 

An infinite hierarchy of layering transitions exists for model polymers in
solution under poor solvent or low temperatures and near an attractive
surface. A flat histogram stochastic growth
algorithm known as FlatPERM has been used on a self- and surface
interacting self-avoiding walk model for lengths up to 256. The
associated phases exist as stable equilibria for large though not
infinite length polymers and break the conjectured Surface Attached
Globule phase into a series of phases where a polymer exists in
specified layer close to a surface. We provide a scaling theory for
these phases and the first-order transitions between them.

\end{abstract}

\maketitle
With the advent of sophisticated experimental techniques
\cite{strick2001a-a}, such as optical tweezers, to probe the behaviour
of single polymer molecules and the explosion of interest in the
physics of biomolecules such as DNA there is a new focus on the study
of dilute solutions of long chain molecules. It is therefore
appropriate to ask whether the thermodynamic behaviour of such
long chain molecules is well understood over a wide range of solvent
types, temperatures and surface conditions.  Even if one considers a
fairly simple lattice model consisting of a self-avoiding walk on a
cubic lattice with nearest-neighbour self-interactions in a half-space
with the addition of surface attraction, the phase diagram has not been
fully explored. In this letter we examine the \emph{whole} phase
diagram highlighting a surprising new phenomenon. In particular, we
demonstrate new features for large but finite polymer lengths
involving the existence of a series of layering transitions at low
temperatures.

Separately, the self-attraction of different parts of the same polymer
and the attraction to a surface mediate the two most fundamental phase
transitions in the study of isolated polymers in solution: collapse
and adsorption. Without a surface, an isolated polymer undergoes a
collapse or coil-globule transition \cite{gennes1972a-a} from a high
temperature (good solvent) state, where in the infinite length limit
the polymer behaves as a fractal with dimension $d_f=1/\nu$ where
$\nu\approx 0.5874(2)$ (known as the \emph{extended} phase) to a low
temperature state (\emph{collapsed}) where the polymer behaves as a
dense liquid drop (hence a three-dimensional globule). In between
these states is the well-known $\theta$-point. Alternately a polymer
in the presence of a sticky wall will bind (adsorb) onto the surface
as the temperature is lowered \cite{debell1993a-a}. At high
temperatures only a finite number of monomers lie in the surface
(\emph{desorbed}) regardless of length even if the polymer is tethered
onto the surface, while at low temperatures a finite fraction of the
monomers will be
\emph{adsorbed} onto the surface in the large length limit and the
polymer behaves in a two-dimensional fashion (with a smaller fractal
dimension of $4/3$). Much theoretical and experimental work has gone
into elucidating these transitions. The situation when both of these
effects are at work simultaneously and hence compete has received
attention in the past decade
\cite{veal1991a-a,vrbova1996a-a,vrbova1998a-a,vrbova1999a-a}. In three
dimensions, four phases were initially proposed: Desorbed-Extended
(DE), Desorbed-Collapsed (DC), Adsorbed-Extended (AE) and
Adsorbed-Collapsed (AC) phases. In the AC phase the polymer is
absorbed onto the surface and behaves as a two-dimensional liquid
drop. Recently a new low-temperature (surface) phase named
Surface-Attached Globule (SAG) \cite{singh2001a-a,rajesh2002a-a} has
been conjectured from short exact enumeration studies and the analysis
of directed walk models \cite{mishra2003a-a}. In this phase the
polymer would behave as a three-dimensional globule but stay
relatively close to the surface. In fact the claim is that there is
not a bulk phase transition between DC and SAG (if SAG exists) in that
the free energy of SAG and DC are the same. However, the number of
surface monomers would scale as $n^{2/3}$, where $n$ is the number of
monomers, rather than the $n^0$ as normally occurs in the desorbed
state \cite{debell1993a-a}.

To explore the phase diagram it is natural to conduct Monte Carlo
simulations. However the scale of the endeavour becomes clear for even
small system sizes (polymer lengths) because one is required to scan
the entire two-energy parameter space of self-attraction and surface
attraction. Even if once fixes one of the parameters, the study of the
properties of the model as the other parameter is varied usually
requires many simulations. Fortunately a recently developed algorithm
\cite{prellberg2004a-a}, FlatPERM, is able to collect the necessary
data in a \emph{single} simulation. The power of this approach cannot
be underestimated. We have utilised FlatPERM to simulate a
self-avoiding walk model of a polymer with both self-attraction and
surface attraction that allows us to calculate quantities of interest
at essentially \emph{any} values of the energies. This was done with
one very long simulation run for polymer lengths up to length
$n_{max}=128$ and also in multiple shorter CPU time runs, up to
polymer length $n_{max}=256$.  We used a Beowulf-cluster to run these
multiple simulations simultaneously with different random number
`seeds'. This allowed us to gain some estimate of statistical errors.

The model \cite{vrbova1996a-a} considered is a self-avoiding walk in a
three-dimensional cubic lattice in a half-space interacting via a
nearest-nearest energy of attraction $\varepsilon_b$ per
monomer-monomer \emph{contact}. The self-avoiding walk is attached at one
end to the boundary of the half-space with surface energy per monomer
of $\varepsilon_s$ for \emph{visits} to the interface. The total
energy of a configuration $\varphi_n$ of length $n$ is given by
\begin{equation}
E_n(\varphi_n) = -m_b(\varphi_n) \varepsilon_b - m_s(\varphi_n)
\varepsilon_s
\end{equation}
and depends on the number of non-consecutive nearest-neighbour pairs
(contacts) along the walk $m_b$ and the number of visits to the planar
surface $m_s$. For convenience, we define $\beta_b =
\varepsilon_b/k_BT$ and $\beta_s =
\varepsilon_s/k_BT$ for temperature $T$ and Boltzmann constant $k_B$.
The partition function is given by $Z_n(\beta_b,\beta_s)=
\sum_{m_b,m_s} C_{n,m_b,m_s} e^{\beta_b m_b+\beta_s m_s}$ with
$C_{n,m_b,m_s}$ being the density of states. It is this density of
states that is estimated directly by the FlatPERM simulation. Our
algorithm grows a walk monomer-by-monomer starting on the surface. We
obtain data for each value of $n$ up to $n_{max}$, and all permissible
values of $m_b$ and $m_s$. The growth is chosen to produce
approximately equal numbers of samples for each tuple of
$(n,m_b,m_s)$. The equal number of samples is maintained by pruning
and enrichment \cite{prellberg2004a-a}. For each configuration we have
also calculated the average height above the surface. Instead of
relying on the traditional specific heat we have instead calculated,
for a range of values of $\beta_b$ and $\beta_s$, the matrix of second
derivatives of $\log(Z_n(\beta_b,\beta_s))$ with respect to $\beta_b$
and $\beta_s$ and from that calculated the two eigenvalues of this
matrix. This gives a clear picture of the phase diagram and allows for
the accurate determination of the multicritical point
\cite{krawczyk2004=c-:a} that exists in the phase diagram (see Figure
1(a)).

We begin our discussion by showing a plot obtained from one run of the
FlatPERM algorithm for $n_{max}=128$ in the region of parameter space
that has been investigated in previous works, namely $0\leq
\beta_b\leq 1.4$ and $0\leq \beta_s\leq 1.6$ (see Figure 1(a)). 
The phase boundaries seen by Vrbov\'{a} and Whittington
\cite{vrbova1998a-a} are clearly visible with four phases in
existence. For small $\beta_b$ and $\beta_s$ the polymer is in a
desorbed and expanded phase (DE). For larger $\beta_s$ adsorption
occurs into the AE phase while for larger $\beta_b$ a collapse
transition occurs into a phase described as the DC by Vrbov\'{a} and
Whittington \cite{vrbova1998a-a} and either SAG or DC by Singh, Giri
and Kumar \cite{singh2001a-a}. We see little evidence for two phases
in this region but given that the SAG/DC phase boundary is not a bulk
phase transition this is not totally surprising!

\begin{figure}[ht]
\begin{center}
\begin{tabular}{cc}
(a) & \\
(b) & \includegraphics[width=86mm]{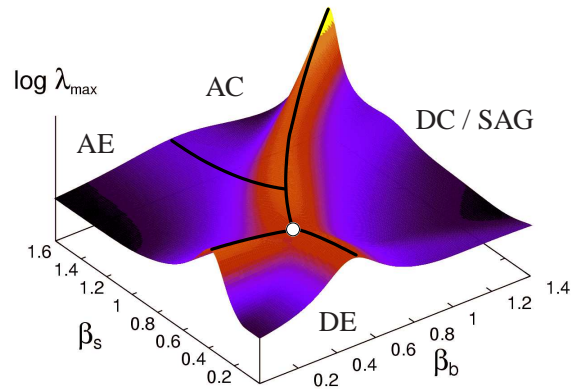}\\
& \includegraphics[width=86mm]{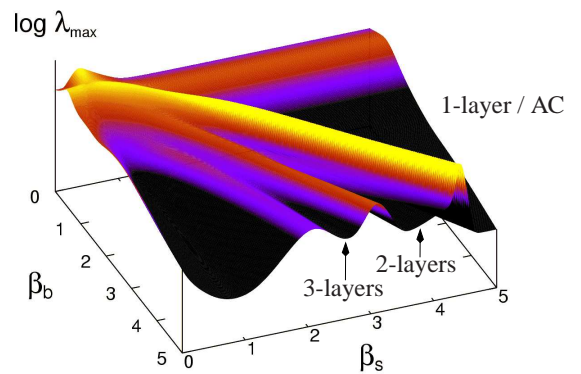}
\end{tabular}
\vspace{3mm}

\caption{\it Plots of the logarithm of the largest eigenvalue of the matrix of
second derivatives of the free energy with respect to $\beta_b$ and
$\beta_s$ for a range of bulk and surface energies.  The lighter the
shade the larger the value. The first figure shows the range of
energies previously considered, and a schematic phase diagram
consistent with Vrbov\'{a} and Whittington.  The white circle denotes
the multicritical point. The second shows an extended range which
clearly shows a new phenomenon (see main text).  These plots were
produced from our $n\leq 128$ long simulation run, using the $n=128$
length data.}

\end{center}
\end{figure}
The power of the FlatPERM method is that it allows us to explore
regions of parameter space usually unavailable to canonical approaches
and so we can consider a much wider range of $\beta_b$ and
$\beta_s$. Of course the price paid is that the polymer lengths
attainable are restricted due to both computer memory required and
time needed to produce the samples.  In Figure 1(b) we consider $0\leq
\beta_b\leq 5.0$ and $0\leq \beta_s\leq 5.0$.

\begin{figure}[ht!]
\begin{center}
  \psfrag{betas}{$\beta_s$}
  \psfrag{ms}{$\langle m_s \rangle / n$}
  \includegraphics[width=76mm]{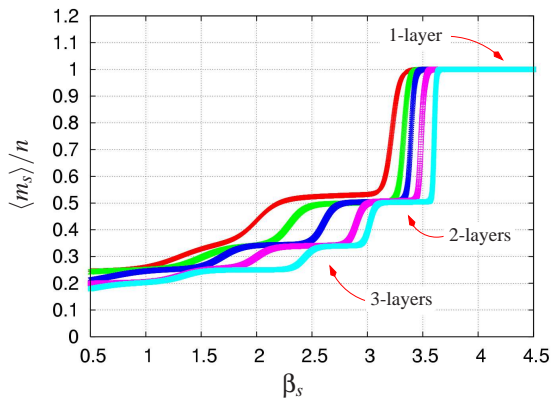}
\vspace{1mm}
\caption{\it A plot of the mean density of visits $\langle m_s \rangle/n$ versus
$\beta_s$ at $\beta_b=4.0$ for lengths $64, 91, 128, 181, 256$ (left
to right) with statistical error estimated as shown.}
\label{coverage}
\end{center}
\end{figure}

To understand what is going on let us consider the mean density of
surface contacts $\langle m_s \rangle/n$ (\emph{coverage}) as a
function of $\beta_s$ at fixed $\beta_b=4.0$ for various values of $n$
up to $n_{max}=256$ (Figure \ref{coverage}). For small $\beta_s$ the
coverage is a slowly varying function of $\beta_s$ and stays that way
as $n$ increases. For $\beta_s$ larger than approximately $\beta_b$ the
coverage converges to a plateau of $1$. So for $\beta_s>\beta_b$
essentially all the monomers are in the surface and the polymer should
behave in a two-dimensional fashion. The transition to the maximum
coverage regime (fully adsorbed) is quite sharp and reflects a first
order phase transition in the thermodynamic limit: we shall confirm
this inference below. The new phenomenon concerns intermediate values
of $\beta_s$ where other plateaus form at around a coverage of $1/2$
and, for larger $n$, also at $1/3$. The transition from one plateau to
another moves towards $\beta_b$ as $n$ increases and also becomes
sharp. We can interpret these intermediate ``phases'' as situations
where the polymer is distributed roughly equally amongst a number of
layers. For example when the coverage is $1/2$ the polymer exists
equally on the surface and in the layer one unit above the surface. As
$n$ increases more and more \emph{layer phases} appear where the
polymer exists in the first $\ell$ layers above the substrate.

\begin{figure}[ht!]
\begin{center}
  \psfrag{betas}{$\beta_s$}
  \psfrag{layers}{$\langle \ell \rangle$}
  \psfrag{height}{$\langle h \rangle / n$}
\includegraphics[width=76mm]{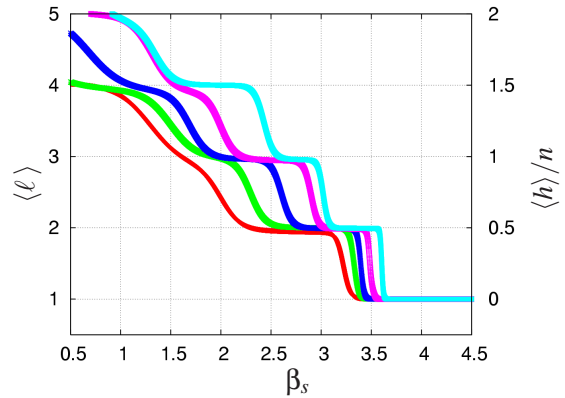}
\vspace{1mm}
\caption{\it A plot of the average height of the polymer per monomer $\langle
h\rangle/n$ (right axis) versus $\beta_s$ at $\beta_b=4.0$ for lengths
$64, 91, 128, 181, 256$ (left to right) with statistical error
estimated as shown. On the left vertical axis the corresponding
average layer number $\langle \ell\rangle$ (assuming uniform
density). }
\label{height}
\end{center}
\end{figure}

To confirm this picture let us consider the mean height of monomers
above the surface $\langle h \rangle$ in Figure \ref{height} for the
same value of $\beta_b=4.0$.
Assuming a uniform density across layers, the mean number of layers
that the polymer subtends, $\langle \ell
\rangle=2\langle h \rangle +1$, can be deduced. We have also estimated 
the end-point position and the maximum height of the polymer, and the
data agrees very well with this assumption. The average height can be
seen to decrease as $\beta_s$ is increased in a series of plateaux
corresponding to the plateaux of coverage. For the range of $\beta_s$
where the coverage is approximately $1/2$ the average height is almost
exactly $0.5$ and the maximum height of a monomer is $1$ (two
layers). Hence the average number of layers is 2, just as our
hypothesis predicts.

To explain the phenomenon of the layering seen above let us examine
the zero-temperature situation.  For positive self-attraction and
surface attraction the polymer will take on some compact configuration
touching the boundary. Consider a Hamiltonian (fully compact)
configuration of fixed height $\ell$ tethered to the surface. In
particular, consider a rectangular parallelepiped with square
cross-section parallel to the surface of side-length $w$. Hence we
have $n=\ell w^2$ and the total energy $E_{\ell}$ (ignoring
contributions from edges and corners) for a $\ell$-layer configuration
is
\begin{equation}
E_{\ell} (\varepsilon_b,\varepsilon_s) \sim - 2\varepsilon_b n +
(\varepsilon_b -
\varepsilon_s) \frac{n}{\ell} +  2\varepsilon_b  \sqrt{\ell n}
\end{equation}
The energy can be minimised for fixed $n$ when
\begin{equation}
 \ell^{3/2}  = \left(1- \frac{\varepsilon_s}{\varepsilon_b}
\right) n^{1/2}
\label{optimumlayer}
\end{equation}
Since the system can only have integer values of $\ell$, a particular
integer value of $\ell\geq 2$ will be stable for a range of
$\varepsilon_s$ of size $O(1/\sqrt{n})$. The AC phase, using this
argument, which is given by $\ell=1$, is stable for
$\varepsilon_s\geq\varepsilon_b$ and for some values of
$\varepsilon_s\leq\varepsilon_b$ given by the relation
(\ref{optimumlayer}). As $\varepsilon_s$ is increased at fixed
$\varepsilon_b$ the system's energy is minimised by smaller values of
$\ell$.  At a fixed value of $\varepsilon_s$ the difference between
the energies of $(\ell+1)$-layers and $\ell$-layers scale as
$(\varepsilon_s - \varepsilon_b) n\ell^{-1}(\ell+1)^{-1}$.
%
It can be argued that non-uniform layers are not stable (consider the
total surface area of a block of smaller width on top of an
$\ell$-layer system) so that the system jumps from $(\ell+1)$-layers
to $\ell$-layers at some value of $\varepsilon_s$.  Hence we deduce
that when the system swaps from $(\ell+1)$-layers to $\ell$-layers
there will be a jump in the internal energy. We expect that this will
be rounded by entropic effects at finite temperatures.

The relation (\ref{optimumlayer}) based on the zero-temperature
energy argument predicts that the transitions coalesce at $\beta_s =
\beta_b$ as $n$ tends to $\infty$. For finite temperatures the
position of the transition need not be exactly $\beta_b$. Let us
denote the infinite $n$ limit transition as occurring at
$\beta_s=\beta^{a}_s$ The thermodynamic limit will realise a sharp
bulk first-order phase transition at $\beta_s = \beta^{a}_s$. At
finite polymer lengths each of the layering transitions are rounded
versions of the zero-temperature jumps in the internal energy. Hence
the layering transitions should be rounded first-order type transition
with specific heat that grows linearly with system size and a
transition width that is $O(1/n)$. Note that since the layer phases
are stable for segments of the $\beta_s$ line of the order of
$1/\sqrt{n}$ and the transitions take place in a region of $\beta_s$
of the order of $1/n$, the transitions become sharper as $n$
increases. Finally, if we consider sitting at fixed value of $\beta_s$
close to $\beta_b$ and increase the polymer length, we should see a
set of layering transitions between phases of layer $\ell$ and
$\ell+1$ with $\ell
\sim n^{1/3}$.

\begin{figure}[ht!]
\begin{center}
\psfrag{yaxis}{$\log(\sigma(m_s)^2/n^2)$}
\psfrag{xaxis}{$ (\beta^a_s - \beta)\sqrt{n}$}
\includegraphics[width=72mm]{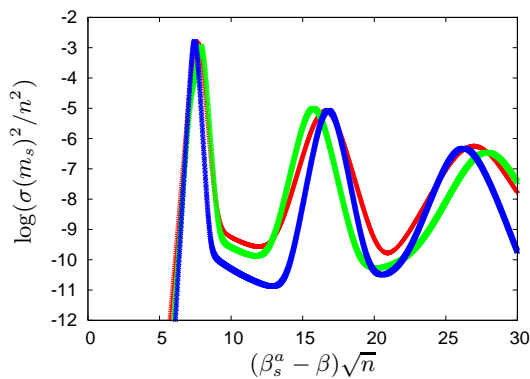}
\caption{\it A plot of  the logarithm of the fluctuations per monomer
divided by $n$ in the number of surface contacts at $\beta_b=4.0$
with the horizontal axis scaled as $(\beta^{a}_{s}-
\beta_s) \sqrt{n} $.  We have used $\beta^{a}_{s}=4.07$. 
Shown are lengths $128, 181, 256$.}
\label{scale-dom}
\end{center}
\end{figure}

We have tested the conclusions of the above argument. At $\beta_b=4.0$
we have estimated the position of the thermodynamic limit transition
to the AC phase (ie 1-layer phase) to be $\beta^{a}_{s}\approx 4.07$
by extrapolating the peaks of the fluctuations in the number of
surface contacts for the strongest transition (ie from 2-layers to
1-layer) against $1/\sqrt{n}$. Figure \ref{scale-dom} shows a scaling
plot of the logarithm of the fluctuations per monomer divided by $n$
in the number of surface contacts against $(\beta^{a}_{s}-
\beta_s)\sqrt{n}$. (For convenience we use the logarithm to display
the needed scale.)  The range includes the peaks from the 1-layer to
2-layers transition, the 2-layers to 3-layers transition and the
3-layers to 4-layers transition.  This demonstrates the scaling
collapse of the height and shift of the layering transitions from
$\ell$-layers to $\ell+1$-layers as $\beta_s$ is decreased.  The
shifts of the peaks of all three transitions scale towards the same
estimate of $\beta_s^a$ when using this same scale, $1/\sqrt{n}$.
The width of the transitions can also be shown to scale with $1/n$,
re-enforcing the hypothesis of first order transitions.

In this letter we demonstrate that the fundamental model of collapsing
and adsorbing polymers in three dimensions contains a new phenomenon
at low-temperatures; at finite polymer lengths a series of (rounded)
layering transitions exist. These transitions increase in number and
become sharper as the polymer length increases.  We note that while
this model is a lattice model, low temperature layering transitions
have been seen in off-lattice models
\cite{celestini2004=a-a}  and arise due to the types of compact
configurations that can occur in the idealised or physical
polymer. The ability to coat a surface with a fixed thickness of
polymer may have experimental and technological applications. We
provide a theoretical framework based on zero-temperature energy
arguments which explain these transitions. The arguments predict that
the transitions coalesce in the infinite length limit to leave a
transition between a collapsed, but not macroscopically adsorbed,
polymer and a collapsed polymer which is fully adsorbed.

\section*{Acknowledgements} 

Financial support from the DFG is gratefully acknowledged by JK and
TP. Financial support from the Australian Research Council is
gratefully acknowledged by ALO and AR. ALO also thanks the Institut
f\"ur Theoretische Physik at the Technische Universit\"at Clausthal.

%
%
%

\end{document}